      [1999/12/01 v1.4c Il Nuovo Cimento]
\documentclass{cimento}

\def\bea{\begin{eqnarray}}
\def\eea{\end{eqnarray}}
\def\beq{\begin{equation}}
\def\eeq{\end{equation}}
\def\bq{\begin{quote}}
\def\eq{\end{quote}}
\def\be{\begin{equation}}
\def\ee{\end{equation}}
\def\bc{\begin{center}}
\def\ec{\end{center}}
\def\bea{\begin{eqnarray}}
\def\eea{\end{eqnarray}}

\def\gappeq{\mathrel{\rlap {\raise.5ex\hbox{$>$}} {\lower.5ex\hbox{$\sim$}}}}
\def\lappeq{\mathrel{\rlap{\raise.5ex\hbox{$<$}} {\lower.5ex\hbox{$\sim$}}}}

\usepackage{graphicx}  

\title{The Early Days of QCD (as seen from Rome)}
\author{G.~Altarelli\from{ins:x}}
\instlist{\inst{ins:x} 
Dipartimento di Fisica `E.~Amaldi', Universit\`a di Roma Tre
and INFN, Sezione di Roma Tre, I-00146 Rome, Italy and \\ CERN, Department of Physics, Theory Unit, 
 CH--1211 Geneva 23, Switzerland}

\PACSes{
\PACSit{11.15,12.38}{\ldots}}
\begin{document}

\maketitle

\begin{abstract}
In honour of Mario Greco I present my recollections on the QCD studies in Rome in the '70's and early 80's and on our very friendly group of people involved.
\end{abstract}

\begin{flushright}
{RM3-TH/11-03}~~~~~~
{CERN-PH-TH/2011-140}\\
\end{flushright}

I have a half-century-long friendship with Mario. We met when students at the University of Rome 
in the early Ô60Õs. Then in '64 I went to Florence and then in '68 to the USA.
When back in Rome in '70 we came in closer contact, also with our families. In fig. 1 one can get an idea of how different we looked at the time. At present we are both at Roma Tre and our offices are a few meters away. In the early '70's QCD and the physics of hard processes was an area of common interest for many of us in Rome and a number of good results were obtained by the different members of our group. Here I will review these results and try to convey the collaborative atmosphere in the group that, for example, led to different collaborations among us to be formed to work on related problems. Of course, while I will talk of the Rome group and of its activity, I am well aware that much more important work on QCD was done at the time in the world, so that I stress that this is not an essay on the history of QCD, but simply a recollection on QCD studies in Rome and on the group of people involved. For a review of the development of QCD in the '70's one can go back, for example, to my '82 review of the subject \cite{GAPR}. 

\begin{figure}
\centering
\includegraphics [width=10.0 cm]{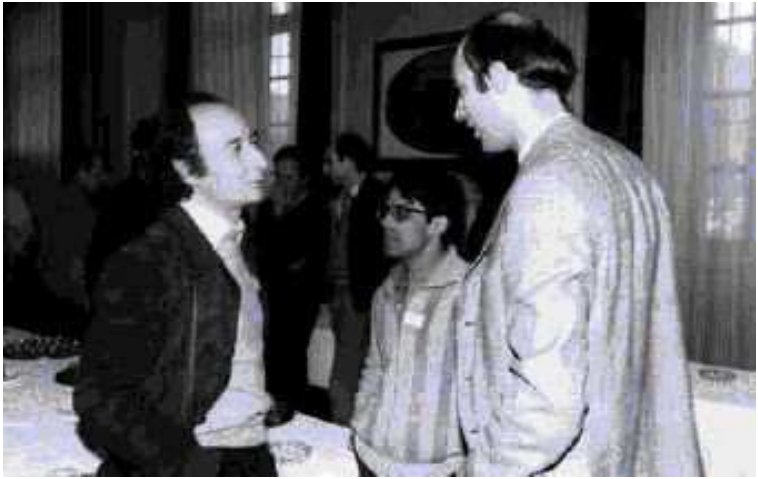}    
\caption{From the left: Mario Greco, Yogi Srivastava and Guido
Altarelli in 1979 at the Accademia dei Lincei, Rome}
\end{figure}

One can argue that QCD really started being a part of the Standard Model with the Nobel-Prize-winning papers by Gross and Wilczek \cite{Gro} and by Politzer \cite{Pol} in '73. Sure enough in previous years there have been very important ground-breaking theoretical works, like those on quarks \cite{qua}, on the naive parton model \cite{part}, on QCD field theory \cite{qcd}, on the renormalization group \cite{ren}  and the short distance operator expansion \cite{opex}. Also Khriplovich, first, and 't Hooft (according to Symanzik), after, apparently discovered asymptotic freedom but did not jump on it as the key property for a theory of strong interactions. In fact the systematic application of QCD to physics only started in ~Õ73. 

At that time three groups were active in Rome, located at Roma 1 ÒLa SapienzaÓ, the Istituto Superiore di SanitaÕ and the Laboratori Nazionali di Frascati
(today in Rome there are three Universities while only "La Sapienza" existed at that time: Roma 2 ÒTor VergataÓ was born in Ô82 and Roma Tre in Ô92). I was in Roma 1 with Nicola Cabibbo and Roberto Petronzio (in '73 he was a student working at his thesis).  Luciano Maiani was at the Istituto Superiore di SanitaÕ, a short walk away from Roma 1, so that Luciano was with us all the time, and Mario Greco with Lia Pancheri, Giorgio Parisi and Yogi Srivastava were at the Laboratori Nazionali di Frascati. Later, important additions to the QCD group in Rome were achieved when Giuseppe Curci, Keith Ellis and Guido Martinelli joined our team.

Even before '73 our group of people  was working in the domain of hard processes and the parton model.  Giorgio Parisi was already known and influential in the study of the physical implications of anomalous dimensions (Giorgio was very young at the time but Kurt Symanzik, a leader in the field, had already a great consideration of him). In a  paper completed in '72 \cite{Par} Giorgio studied the deep inelastic scattering structure functions in a $\lambda \phi^4$ theory with negative coupling $\lambda$ < 0, a theory discussed by Symanzik as a field theory model for Bjorken scaling. This paper was cited by Gross and Wilczek \cite{Gro} and also in the review by Gross, "Asymptotic Freedom and QCD - a Historical Perspective" \cite{Gro2}, written shortly after he got the Nobel Prize. In another '72 paper \cite{Par2} Giorgio derived limits on logarithmic scaling violations in deep inelastic scattering structure functions from the existing data. This work was quoted in the asymptotic freedom paper by Politzer \cite{Pol}. In Ô73 at "La Sapienza" we were studying hard processes in the parton model (with scaling). Luciano Maiani and myself we studied deep inelastic processes in the $\lambda \phi^3$ theory, a superrinormalisable model for scaling \cite{phi3}, in continuation of previous work \cite{Rub}. With Cabibbo and Petronzio the two of us completed in '73 a series of papers on the nucleon as a bound state of 3 quarks \cite{ACMP}, where a parton picture of the structure functions was developed, with the nucleon described in terms of constituent quarks, each of them with a parton structure (so that for example the proton and the pion structure functions could be related through the constituent structure functions). This idea is still viable and it is in competition with the picture of the nucleon as 3 valence quarks floating in a sea of quarks and gluon partons (with no separation of the 3 constituent quarks). In Ô72 Mario Greco, with Bramon and Etim, evaluated the hadronic contribution to the muon anomalous magnetic moment from the data on $e^+e^-$ cross-sections \cite{Gre1}. This is a  problem of high current interest still today. Mario and collaborators found at the time $a_\mu=68\pm9$ with a linear sum of errors, which corresponds to $a_\mu=68\pm6$ with errors summed in quadrature. This value is to be compared with the modern estimate $a_\mu=69.23\pm0.42$ by Davier et al \cite{Dav}. At that time Mario was interested in hard processes and was advocating a model for $e^+e^-$ annihilation, deep inelastic scattering and Drell-Yan processes based on extended vector boson dominance \cite{Gre2}. The functional behaviour of the couplings versus mass of the tower of vector bosons was chosen as to get approximate scaling. This approach was a competitive picture with respect to the parton model for some time but it was later abandoned because it predicted no jets in $e^+e^-$ annihilation  and no suppressed ratio of longitudinal over transverse cross-sections in deep inelastic scattering.

After the Gross-Wilczek and Politzer papers we immediately turned to study the potentiality of QCD for improving the parton model. Myself and Maiani we decided to study the QCD corrections to the effective weak non-leptonic Hamiltonian, written  as a Wilson expansion in terms of 4-quark operators of the (V-A)x(V-A) type obtained by integrating away the $W^\pm$ exchange \cite{AM}. The logarithmically enhanced terms of the QCD corrections are fixed by the anomalous dimensions of these operators, much in the same way as the moments of structure functions get logarithmic corrections as computed by Gross et al \cite{Gro, Pol} from the anomalous dimensions of the leading-twist operators in the light-cone expansion. Our hope was to find that the QCD corrections act in the direction of enhancing the $\Delta T=1/2$ operators with respect to those with $\Delta T=3/2$, thus explaining, at least in part, the empirical $\Delta T=1/2$ rule (where $T$ is the isotopic spin). The explicit calculation turned out to lead to precisely this result, as also obtained in a simultaneous work by M. K. Gaillard and B. W. Lee \cite{Gai} (actually these authors had pointed out to us the crucial role of charm in this problem). These important papers were the first calculations of the QCD corrections to the coefficients of the Wilson expansion in the product of two weak currents, an approach that, suitably generalised (by considering other weak processes) and improved (for example, by computing the anomalous dimensions beyond the leading order), still represents a basic tool in this field. In the following months we applied the method to charm decays  \cite{ch}, before the discovery of charm, and to weak neutral current processes \cite{nc}. To this last paper also contributed Keith Ellis, a scottish PhD student of Cabibbo, who was to stay with us in Rome for a few years, eventually speaking a very good italian and fully understanding the roman way of living. Later, in '81 myself with Curci (who, unfortunately, is no more with us), Martinelli and Petrarca \cite{2loop} we computed the two-loop anomalous dimensions for the operators of the effective weak non-leptonic Hamiltonian.

Meanwhile Mario Greco, working in Frascati with Touschek (and Pancheri, Srivastava and Etim) was becoming an expert in QED radiative corrections and the resummation of soft photons. In fact, the $e^+e^-$ collider ADONE was  functioning at the time and this prompted QED studies as a main activity of the theory group of the Laboratory.  For example, in '75 Mario published two papers on the QED corrections near the $J/\Psi$ \cite{GreJ}. This work has been later generalized to the production of the Z boson, preparing the stage for the  analysis of  LEP/SLC experiments \cite{GreZ}. With the advent of QCD he could profit of the acquired expertise in QED naturally turning into resumming soft gluons (see later). I remember that I learned from Mario's papers the techniques and the results of the exponentiation of logs in QED. 

In '75 Cabibbo and Parisi published an important work. This is one of the first papers where quark deconfinement is discussed \cite{Cab}. In this work they argue that an exponentially growing hadronic spectrum (aÕ la Hagedorn), which is compatible with the ever increasing population of observed hadronic resonances, can be naturally associated to a 2nd order phase transition that one could identify with the deconfining transition from the hadronic phase into that corresponding to the quark-gluon plasma. 

At about the same time myself, in collaboration with Parisi and Petronzio, we studied the QCD corrections to neutrino deep inelastic cross-sections and distributions \cite{yan}.  We found that the corrections, also including those due to the onsetting of the charm threshold, are rather large at the energy of the then available experimental data. In the absence of these corrections the data appeared at variance with respect to the predictions of the parton model. This paper contributed to the downgrading of the observed so called Òy-anomalyÓ from a signal of new physics (right-handed charged currents were invoked) down to a less exciting charm threshold plus QCD-logs effect.

In '77 the well known work on the QCD evolution equations by myself and Parisi was published \cite{AP}. In the academic year '76-77 both of us were on sabbatical in Paris. I was at the Ecole Normale Superieure (ENS) and Giorgio at the Institut des Hautes Etudes Scientifiques at Bures-sur-Yvette. Often Giorgio preferred to stay downtown, spending some time at the ENS to discuss with the people there and, in particular, with me on QCD phenomenology, a subject of great interest for both of us at that time. Out of these regular contacts our work on the evolution equations was developed. The main virtue of our approach was to formulate the evolution of parton densities as a branching process with probabilities determined (at leading order) by the splitting functions (proportional to the running coupling). In our paper a particular emphasis was devoted  to prove that the splitting functions are a property of the theory and do not depend on the process (in particular the evolution does not apply only to deep inelastic scattering). On this issue I remember a discussion some months before with Cabibbo, who was asking what remains of the parton model if the scaling violations modify the parton densities in different ways for different processes. I argued that an appealing possibility was that the leading logarithmic corrections are universal and that general $Q^2$-dependent parton densities could be defined in this limit and used for the description of a variety of hard processes (what is now denoted as the "factorization" theorem). With this idea in mind, in our paper, completely formulated in parton language, with running coupling, the splitting functions were directly derived from the QCD vertices, using the formalism of the "old" perturbation theory (because the 3 partons in the vertex cannot all be on their mass shell), with no reference to the particular diagram where the splitting leg is attached to, thus making clear that the splitting functions are the same for all processes. The polarized splitting functions were also derived by us with the same method in agreement with the results of refs. \cite{pol} obtained by the operator method.

The evolution equations are now often called DGLAP equations (Dokshitzer Gribov Lipatov Altarelli Parisi). The first article by Gribov and Lipatov was published in '72 \cite{Gri} (even before the works by Gross and Wilczek and by Politzer!) and was followed in '74 by a paper by Lipatov \cite{Lip} (these dates correspond to the publication in russian). All these articles refer to an abelian vector theory (treated in parallel with a pseudoscalar theory). Seen from the point of view of the evolution equations, these papers, in the context of the abelian theory, ask the right question and extract the relevant logarithmic terms from the dominant class of diagrams. But from their formal presentation the relation to real physics is somewhat hidden (in this respect the '74 paper by Lipatov makes some progress and explicitly refers to the parton model). The article by Dokshitser \cite{Dok} was exactly contemporary to ours. It now refers to the non abelian theory (with running coupling) and the discussion is more complete and explicit than in the Gribov-Lipatov articles. But, for example, the notion of the evolution as a branching process and the independence of the kernels from the process are not emphasised.  An important point is also that the Gribov-Lipatov papers were known to Dokshitser (while not to us). Their works were in fact his starting point and are quoted among the references given in his article.

Back to Rome I met Guido Martinelli, then a post-doc with a contract for doing accelerator physics at Frascati, and I rescued him into particle physics, with a work on the transverse momentum distributions for jets in lepto-production final states \cite{Mart}. In the same paper we derived an elegant formula for the longitudinal structure function $F_L$, also an effect of order $\alpha_s(Q^2)$, as a convolution integral over $F_2(x,Q^2)$ and the gluon density $g(x,Q^2)$. I find it surprising that it took ~40 years since the start of deep inelastic scattering experiments to get meaningful data on the longitudinal structure function. The present data, recently obtained by the H1 experiment at DESY, are in agreement with this LO QCD prediction but the accuracy of the test is still far from being satisfactory for such a basic quantity.

Meanwhile Mario Greco started producing an impressive series of works where the tools developed over the years for QED were applied to the resummation of soft gluons in different QCD processes. In a first group of 
papers \cite{coh}, \cite{CGS} the QED formalism of coherent states was adapted to the non abelian context of QCD. In particular this technique was applied by Mario with Curci and Srivastava \cite{CGS} to compute the probability that a fraction $\epsilon$ of the total energy $2E$ falls outside a cone of semi-aperture $\delta$. This amounts to upgrading the Sterman-Weinberg perturbative result obtained in Ô77 \cite{sw} by including soft gluon resummed effects.

\begin{figure}
\centering
\includegraphics [width=10.0 cm]{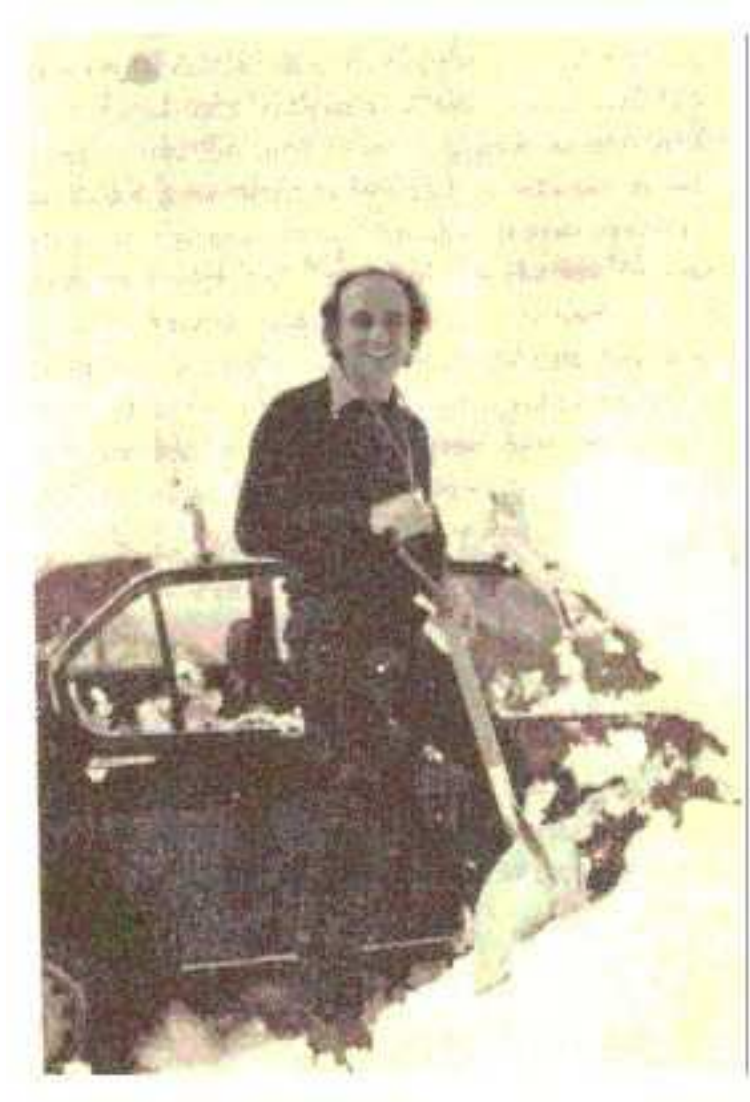}    
\caption{Mario Greco at a Moriond meeting in 1980 where he gave
the talk ÒSoft gluon effects in QCD processesÓ}
\end{figure}

Resummation near the phase space boundaries is an important issue in QCD. Mario authored with Curci one of the early papers on this subject \cite{GCu}, with applications to deep inelastic scattering near $x=1$ and to Drell-Yan processes near $\tau=Q^2/s=1$. The resummation of the "large $\pi^2$ terms" (those arising from the continuation of $Q^2$ from negative values in deep inelastic scattering, where the parton densities are measured, to the positive values of the Drell-Yan process), was also included in this paper (Parisi also studied this problem at near the same time \cite{ParPi}). In fig. 2 we can see Mario at work during a Moriond meeting where he presented those results.

In those years the Rome group contributed very much to the theory of Drell-Yan processes. In addition to the works just mentioned, important progress was made in '78-'79 with the calculation of the next to the leading (NLO) corrections to Drell-Yan processes by myself with Keith Ellis and Martinelli \cite{AEM}. This was one of the first calculations of NLO corrections in QCD. We started by defining the quark parton densities beyond leading order in a precise way (for quarks we adopted the structure function $F_2$ as the defining quantity: the naive parton model expression is taken by definition to hold unchanged at NLO; gluons only enter at NLO in Drell-Yan processes). Then the calculation of NLO diagrams for both deep inelatic scattering and the Drell-Yan process allows to derive the corrective terms for the Drell-Yan cross-section, as function of $Q^2$. The resulting corrections turned out to be surprisingly large. The ratio of corrected to uncorrected (Born) cross-sections was found to be rather constant in $Q^2$ and in rapidity. So we decided to denote it as the "K-factor", because K sounded to us as the typical symbol for a constant. The origin of the main part of this correction can be traced back to effects that can be resummed (like the "large $\pi^2$ terms" that we have just mentioned).  Today with much larger values of $Q^2$ and $s$ accessible to present accelerators and with the NNLO calculations completed the $K$ factor is under control (for example, there is not too much difference between NLO and NNLO estimates, especially when some resummations are also implemented)

Another by now classic theoretical problem for Drell-Yan processes that was first attacked in those years is the evaluation of the transverse momentum ($p_T$) distribution of the produced virtual boson (a $\gamma$ or a $W^\pm$ or a $Z_0$). The study of the LO perturbative $p_T$ distribution, valid for $p_T \sim Q$, was completed in '78 by myself with Parisi and Petronzio \cite{pT1}. The NLO perturbative calculation followed in '81-'83 by K. Ellis, Martinelli and Petronzio \cite{pT2}. The study of the Sudakov double logs, important at intermediate values of $p_T$ (between $\Lambda_{QCD}$ and $Q$) was started in '79 by Mario with Curci and Srivastava \cite{CGS} and by Parisi and Petronzio \cite{sud2} (in this paper the completely correct formula for the LO Sudakov factor was first obtained, correcting a small bug in a previous paper by Dokshitser, Dyakonov and Troyan \cite{sud3}). Then, in the early '80, the problem was  attacked of realizing a smooth matching between the perturbative and the Sudakov component. Mario worked on this problem with Pierre Chiappetta \cite{mat1}. As soon as the data on the $W$ and $Z$ production from $U1$ and $U2$ at CERN were first available, an adequate theoretical prediction was ready in a paper signed by myself, K. Ellis, Greco and Martinelli \cite {mat2}. This is an important paper, first because it is a paper that I signed with Mario, and then because it essentially contained all the crucial ingredients that describe the physics of this phenomenon. In the subsequent years the accuracy was much improved with the computation of subleading effects and with several different refinements, but the essential points were all present in our paper and the accuracy of our treatment was adequate for the quality of the first data. The same techniques are at present applied to the calculation of the $p_T$ distribution of the Higgs boson produced by gluon fusion (see, for example, ref. \cite{Cat}).

I leave to the following speakers to describe other aspects of the scientific activity of Mario and also what he did later in QCD. I stop here by making my best wishes  to him of a long and happy
sequel of celebrations for 75, 80, 85,....100 anniversaries. And also I offer my compliments to him for the
co-foundation of these by now classical meetings in La Thuile. 

\acknowledgments
I thank the Organizers of Les Rencontres, for giving me the opportunity to celebrate the 70th anniversary of my old friend and colleague Mario Greco.

\end{document}